\documentclass[twocolumn,prl,10pt,amsmath,amssymb,nofootinbib,showpacs,superscriptaddress,floatfix]{revtex4-1}

\DeclareFontFamily{U}{rcjhbltx}{}
\DeclareFontShape{U}{rcjhbltx}{m}{n}{<->rcjhbltx}{}
\DeclareSymbolFont{hebrewletters}{U}{rcjhbltx}{m}{n}

\usepackage{graphicx}
\usepackage{color}
\usepackage[usenames,dvipsnames]{xcolor}
\usepackage[colorlinks=true,linkcolor=Red,citecolor=Green,linktoc=page]{hyperref}
\usepackage{multirow}
\usepackage{float}
\usepackage{flushend}
\usepackage{balance}
\usepackage[varg]{txfonts}
\usepackage{ulem}
\usepackage{fancyhdr}

\DeclareMathSymbol{\lamed}{\mathord}{hebrewletters}{108}

\begin{document}
\title{Universal upper bound for the entropy of superconducting vortices and the quantum Nernst effect}
\author{M.\,C.\,Diamantini}
\affiliation{NiPS Laboratory, INFN and Dipartimento di Fisica e Geologia, University of Perugia, via A. Pascoli, I-06100 Perugia, Italy}
\author{C.\,A.\,Trugenberger}
\affiliation{SwissScientific Technologies SA, rue du Rhone 59, CH-1204 Geneva, Switzerland}
\author{V.\,M.\,Vinokur}
\affiliation{Terra Quantum AG, St. Gallerstrasse 16A, CH-9400 Rorschach, Switzerland}
\affiliation{City College of the City University of New York,
	160 Convent Ave, New York, NY 10031, USA}

\begin{abstract}
We show that the entropy per quantum vortex per layer in superconductors in external magnetic fields is bounded by the universal value $k_{\rm B}{\rm ln}\,2$, which explains puzzling results of recent experiments on the Nernst effect. 
 \end{abstract}

\maketitle

\section{Introduction}
The Nernst effect\,\cite{nernst,varl2018} is the emergence of a transverse electric field produced by a longitudinal thermal  gradient  in  the presence  of  a magnetic  field and measures the 
flow of transverse entropy induced by a longitudinal particle motion. The Nernst effect  has attracted a great deal of attention 
after the discovery of a sizable Nernst coefficient in high-temperature cuprate superconductors due to fluctuating Cooper pairs and mobile vortices simultaneously carrying entropy and magnetic flux.  The high interest in the Nernst effect is amplified by the fact that measurements of the Nernst signal can provide information about material parameters inaccessible by other means, for example, the upper critical field $H_{\mathrm c2}$, which often cannot be directly measured because of its large value\,\cite{chang}.
The Nernst signal has a maximum as a function of temperature and magnetic field and a tail into the normal state. Recent experiments\,\cite{universal} revealed that, in the fluctuation region, the Nernst effect has the magnitude expected by theory. However, the peak amplitude of the Nernst signal is unexpectedly about the same in different superconductors despite the broad variance in basic superconducting parameters, and it corresponds to the universal value of an entropy per vortex per layer $\approx$$k_{\rm B}\,{\rm ln}\,2$ \cite{universal} (see also \cite{huebener}). 

Here, we demonstrate that this remarkable finding is an immediate consequence of the dynamical symmetry of a superconducting ground state under the algebra \linebreak $\left( W_{1+\infty} \otimes \overline W_{1+\infty} \right)/\widehat U(1)$, where $W_{1+\infty}$ is the algebra of quantum area preserving diffeomorphisms\,\cite{bakas1, bakas2, pope} (see\,\cite{shen} for a review).

%%%%%%%%%%%%%%%%%%%%%%%%%%%%%%%%%%%%%%%%%%
\section{Methods}
Dynamical symmetries are very powerful tools for deriving the structure of a classical configuration space or a quantum Hilbert space and the properties of their excitations once the relevant symmetry governing them are identified, as in the paradigmatic example of the flavor SU(3) symmetry of strong interaction. Uniform ground states with a gap---like the Cooper pair condensate in superconductors---are incompressible in the limit of a large gap, where density waves are suppressed. Note that although in a d-wave high-$T_c$ superconductor, one could expect the gapless nodal modes to promote charge density waves (CDW), the maximum Nernst signal is observed in the region where superconductivity dominates\,\cite{universal} and the competing CDW order does not develop. Hence, Cooper pair condensate incompressibility is preserved there as well. 

The dynamical symmetry of a gapped system (in the limit of the large gap), classifying all possible configurations at the classical level, or states at the quantum level, is that of volume-preserving diffeomorphisms. In the presence of an applied external magnetic field, or for thin films, the dynamical symmetry reduces to area-preserving diffeomorphisms (in the plane orthogonal to the magnetic field if this is present). An area-preserving diffeomorphism transforms, for example, a circular region of the plane into a deformed region of the same area, and it is thus the relevant symmetry for incompressible fluids, for which deformations of the constant area of a droplet are the only allowed transformations. 

The 2D case is particularly interesting since the algebra of area-preserving diffeomorphisms is a well-known extension of the 2D conformal algebra\,\cite{difrancesco}. An example of the volume-preserving transformations should immediately come to mind from classical mechanics: the canonical transformations preserve the volume of phase space. When the system is 1D, the phase space is 2D, and canonical transformations preserve its area. It~turns out that at the classical level, the area-preserving diffeomorphisms can always be represented as canonical transformations of a two-dimensional phase space. In the gapped systems, there is always a fundamental length associated with the gap. For quantum Hall systems, it is the magnetic length $\ell_{\mathrm H}=\sqrt{\hbar c/eB}$, where $\hbar=h/2\pi$, $h$ is the Planck constant, $c$ is the light velocity, $-e$ is the electron charge, and $B$ is the magnetic field. In superconductors, it is $\ell_{\mathrm{\xi}} = O(\xi)$, where $\xi$ is the coherence length. From now on, we normalize all lengths by this fundamental length $\ell_{\mathrm{\xi}}$, endowing the plane with coordinates $z$ and $\bar z$ with a Poisson bracket
\begin{equation}
	\{ f,g \} = \pm i \ ( \partial f \bar \partial g - \bar \partial f \partial g )\ ,
	\label{ab}
\end{equation}
so that $z$ represents a coordinate, and $\bar z$ its conjugate momentum (or the other way around).
One describes the area-preserving diffeomorphisms as canonical transformations $\delta z=\{ {\cal L}, z\} $ and $\delta \bar z=\{ {\cal L}, \bar z \}$ with generating functions ${\cal L}(z, \bar z)$. The basis of generators ${\cal L}_{n,m}=z^n \bar z^m$
satisfies what one calls the classical $w_{\infty}$ algebra \cite{shen}
\begin{equation}
	\left\{ {\cal L}_{n,m}, {\cal L}_{k,l} \right\} = \mp i \ (mk-nl) \ {\cal L}_{n+k-1,m+l-1} \ ,
	\label{ac}
\end{equation}
which is obtained simply by repeatedly using the Poisson bracket (\ref{ab}) on the generators.
The operators ${\cal L}_{nm}$ with $n > 0$ and $m > 0$ form two closed sub-algebras, the two chiral sectors of the classical $w_{\infty}$ algebra, related by complex conjugation. Generators with both $n$ and $m$ being negative are called descendants and are obtained as products of generators in the two fundamental chiral sectors. Finally, generators with $n=0$ and $m=0$ form two Abelian subalgebras. For systems which do not break parity, like the present one, we use different signs in the definition of the Poisson bracket (\ref{ab}) for the two chiral sectors. A parity transformation involves complex conjugation and exchange of the two chiral sectors so that the symmetry algebra (\ref{ac}) is left invariant. 

The quantum version of this infinite-dimensional algebra is obtained by the usual substitution of Poisson brackets by quantum
commutators: $i\{ ,\} \rightarrow [ , ]$. Let us denote the quantum version of ${\cal L}_{i-n,i}$ by $V^i_n$, and let us first discuss one single chiral sector of the quantum algebra by restricting to positive values of $i$. This gives the algebra $W_{\infty}$, 
\begin{eqnarray}
	{[ V^i_n, V^j_m]} = (jn-im) \ V^{i+j-1}_{n+m} +q(i,j,n,m)\ V^{i+j-3}_{n+m}
	+\cdots \nonumber \\+\delta^{ij}\delta_{n+m,0}\ c\ d(i,n) \ ,
	\label{ad}
\end{eqnarray}
where the structure constants $q$ and $d$ are polynomials of their arguments and the dots denote a finite number of
similar terms involving the operators $V^{i+j-1-2k}_{n+m}$. The first term in the r.h.s.\,of Equation\,(\ref{ad}) is the classical term
(\ref{ac}). The remaining terms are quantum operator corrections, with the exception of the last c-number term, which represents a quantum anomaly with the central charge $c$. All the quantum operator corrections are uniquely determined by the closure of the algebra; only the integer central charge $c$ is a free parameter. If we admit also the value $i=0$, we obtain the full algebra $W_{1+\infty}$, including the Abelian subalgebra quantum generators $V^0_n$. The quantization of the full classical algebra $w_{\infty}$, involving generators in the two chiral sectors, the two Abelian subalgebras, and all their products, is then obtained as the direct product of two copies $W_{1+\infty}$ and $\overline W_{1+\infty}$ of opposite chirality. The exact form of the polynomials $q$ and $d$ are not relevant for what follows, and~we thus omit them for simplicity.

%%%%%%%%%%%%%%%%%%%%%%%%%%%%%%%%%%%%%%%%%%
\section{Results}
Let us first consider, for simplicity, a single chiral sector and give some examples. The~generators $V^i_n$ are characterized by an integer conformal (scaling) dimension \linebreak $h=i+1 \ge 1$ and a mode index $n$, $-\infty < n < +\infty $. The operators $V^0_n$ satisfy the Abelian Kac--Moody algebra (see\,\cite{difrancesco} for a review), $\widehat U(1)$, which is the quantum extension of the usual $U(1)$ by a c-number central charge, while the operators $V^1_n$ are the generators of conformal transformations, satisfying the Virasoro algebra\,\cite{difrancesco}
\begin{eqnarray}
	{[V^0_n,V^0_m]}    && =  n \ c\ \delta_{n+m,0} \ ,
	\nonumber \\
	{[ V^1_n, V^0_m }] && =  -m\ V^0_{n+m} \ ,
	\nonumber \\
	{[ V^1_n, V^1_m ]} && =  (n-m)V^1_{n+m} +{c\over 12}n(n^2-1)
	\delta_{n+m,0}\ .
	\label{ae}
\end{eqnarray}

The operators $V^0_n$ and $V^1_n$ are the charge and angular momentum modes in the chiral sector under consideration. 

Exactly as in the familiar representation theory of the rotation group SU(2), a multiplet is a representation of the symmetry algebra. This is formed by the highest weight state, which is annihilated by all lowering operators, for example $S_z =-1$, and further states obtained by applying on it the raising operators, for example $S_z = 0$ and $S_z = +1$. The~only difference here is that the symmetry algebra is infinite-dimensional, and thus there are infinite ``spin" operators labeled by $i$ and infinite lowering and raising operators labeled by $\pm n$. The incompressible quantum ground state is thus a highest-weight state $|\Omega\rangle_W$ satisfying
\begin{eqnarray}
	V^i_n \vert\Omega\rangle_W &&=0\ , \quad \forall\ n >0\ ,\ i\ge 0\ ,
	\nonumber \\
	V^i_0 \vert\Omega\rangle_W &&=0\ ,\ \ \ i\ge 0\ .
	\label{ai}
\end{eqnarray}

The particle-hole excitations are obtained by applying generators with negative mode index (the raising operators) to $\vert\Omega\rangle_W$. Due to incompressibility, these gapless excitations are edge excitations. This is not the only possibility, however.
There can be other highest-weight representations of $W_ {1+\infty}$ for which the operators $V^i_0$ do not vanish. These are identified 
with the possible bulk excitations, with quantum numbers encoded in $V^i_0$ and each with its tower of gapless edge excitations. As in the case of SU(2), the highest-weight representations can be composed. The rules to do so are called fusion rules in this infinite-dimensional case\,\cite{difrancesco}. Finally, there are operators $V(z)$ that interpolate between different highest-weight representations. These operators, called vertex operators, can be thought to insert a given excitation at $z$ when applied on the ground state\,\cite{difrancesco}. 

This representation theory of $W_{1+\infty}$ has been applied to classify incompressible quantum fluids corresponding to the observed hierarchy of quantum Hall plateaus \cite{ctz1, ctz2, ctz3}. In~the case of superconductors \cite{ctsuper}, in which there is no parity-breaking magnetic field, we start with the direct product of two copies of the symmetry algebra of opposite chirality, $W_{1+\infty} \otimes \overline W_{1+\infty}$. However, the charge operators $V^0_n +\overline {V^0}_n$ spanning the charge Kac--Moody algebra $\widehat U(1)_{\rm charge}$ have to be modulated out from this symmetry group since the charge is condensed and does not represent a good quantum number. The dual group $\widehat U(1)_{\rm vortex}$, spanned by the operators $V^0_n - \overline {V^0}_n$, represents vortices with their tower of edge excitations, $V^0_0 - \overline {V^0}_0$ being the vortex number. Actually, this entails that a single axial $W_{1+\infty}$ remains as the full symmetry group, with its $\widehat U(1)$ Kac--Moody algebra encoding the Abelian vortices~\cite{ctsuper}. 

Let us now consider a quantum vortex at holomorphic coordinates $(z, \bar z$) or, equivalently, at $(x= (z+\bar z)/\sqrt{2}, y= (z-\bar z)/i\sqrt{2} )$. This is described by a wave function $\Phi (z, \bar z)$, which is the direct product of two chiral wave functions, $\Phi (z, \bar z) =\psi_{\rm L} (z) \psi_{\rm R} (\bar z)$, each of which corresponds to a highest-weight state in its chiral symmetry sector. We can thus focus on each of these sectors separately. As we have derived, the incompressibility implies that two coordinates $z$ and $\bar z$ corresponding to the generators $V^0_{-1} = z$ and $V^1_1= \bar z$ do not commute as a consequence of the symmetry algebra, 
\begin{equation}
	\left[ z, \bar z \right] = \left[ V^0_{-1}, V^1_1 \right] = 1 \ ,
	\label{notcommute}
\end{equation}
or, equivalently,
\begin{equation}
	\left[ x, y \right] = -i \ ,
	\label{notcommutexy}
\end{equation} 

Incompressibility leads to a phase-space reduction in each chiral sector, the coordinate space itself becomes a phase space, and the two coordinates do not commute. In the presence of the magnetic field, this is tantamount to the well-known non-commutativity of the magnetic translations or of the guiding centres of charged particles in the first Landau level. Finite magnetic translations form nothing else but the group of the area-preserving diffeomorphisms on the torus. Note, however, that an external magnetic field is just one possible mechanism leading to an incompressible quantum fluid. Other mechanisms can exist that lead to the same non-commutativity as a consequence of their dynamical symmetry\,\cite{ds}. In~the case of superconductors, the condensate is the origin of the non-commutativity. While in the magnetic case, the area that appears on the right-hand side of the non-normalized coordinate commutator is the square of the magnetic length, for superconductors it is the square of the superconducting correlation length $\xi$. For small values of the magnetic field, the gap relevant for incompressibility remains the superconducting gap, and the dynamical symmetry is preserved. This happens when the magnetic length is larger than the correlation length, $eB /\hbar c <  1/\xi^2$, which can be rewritten as $B < \Phi_0 /2\pi \xi^2 = B_{\rm cr2}$, with $\Phi_0$ the flux quantum and $B_{\rm cr2}$ the upper critical field of the type II superconductor. 

The immediate consequence of this non-commutativity is that the vortex wave function is a direct product of chiral wave functions 
of only one of the coordinates $x$ and $y$, while the other is realized as a momentum. The choice of which is a true coordinate and which is the conjugate momentum is called a choice of polarization in field theory. For~example, we can choose the vortex wave function as $\Phi (y)$ and realize $x$ as $x=-i d/dy$. Both possibilities are legitimate, and the polarization choice does not influence physical quantities. The really important point is that, as in standard quantum mechanics, due to incompressibility we have a generalized Heisenberg uncertainty relation between true coordinate and momentum, 
\begin{equation}
	\Delta x \Delta y \ge \ell_{\mathrm{\xi}}^2 = O\left( \xi^2 \right) \ ,
	\label{ghr}
\end{equation}
where we have used standard lengths to best expose the physical content of this equation. This means that the position of the quantum vortex in the plane cannot be specified more precisely than the area of $O(\xi^2)$.  As in the standard quantum statistical mechanics, the 2D phase space decomposes into the fundamental cells of ``area'' $\hbar$ with each such cell containing no more than one fundamental quantum degree of freedom, and here the actual plane decomposes into the real cells of the area $A_{\rm fund}= O(\xi^2)$, each capable of accommodating at best one quantum vortex. Note that these cells do not have to be regular; it is only the area which is important. When all the cells are occupied, one cannot squeeze another vortex into the system since this would violate the condition (\ref{ghr}).

Let us consider the possible distribution of quantum vortices. To that end, we divide the sample area $A$ into $N=A/A_{\rm fund}$ cells, which can each accommodate exactly one quantum vortex in a maximally squeezed configuration. An upper bound for the number of possible configurations can be easily found by assuming that each such cell has two possible states: it is either occupied by a vortex or not. This would correspond to the number of available configurations $2^N$ with $N=A/A_{\rm fund}$ the number of fundamental cells fitting in the sample, corresponding to an entropy bound $S= Nk_{\rm B} {\rm ln}\,2$. This bound becomes exact when the number of vortices reaches the maximum value $N$, that is, when all cells are occupied. In a maximally squeezed configuration, each cell is occupied, and each quantum vortex thus carries the universal entropy $k_{\rm B}{\rm ln}\,2$, in full accord with the Nernst effect measurements\,\cite{universal}. Note that we have not used any information other than the presence of a gap in an effective 2D system. The universal bound is simply a consequence of the incompressibility of the superconducting ground state and is common to all possible superconductors, independently of the details of their material parameters.

As soon as the applied magnetic field has created enough quantum vortices to completely fill up the available cells in a given sample, the further vortex accommodation upon further small increase of the magnetic field becomes impossible. As a consequence, superconductivity breaks down. This is thus an alternative characterization of the upper critical field in type II superconductors. The universal entropy limit per vortex is reached precisely at this value of the applied field in a full accord with the experimental observation\,\cite{universal}.

\section{Discussion and Conclusions}

We would like to stress that our derivation for the gapped systems does not preclude the possibility of other, gapless systems having quasiparticles with entropies of $O(k_{\rm B})$. A~gap is a sufficient condition for the universal entropy bound in effective 2D systems but not a necessary one. 

Finally, let us note that this result can be interpreted in terms of information theory. %as a consequence of the celebrated information Landauer bound \cite{landauer}. %The upper bound $k_{\rm B}{\rm ln}\,2$ on the entropy of a vortex can be viewed also as a lower bound on its internal energy. In the Nernst experiment vortices, the fundamental degrees of freedom are ``erased" from the sample when they are measured at the boundaries.  As derived by Landauer, the absolute minimum energy required for such an ``erasure" is exactly $k_{\rm B}{\rm ln}\,2$. 
A maximally squeezed vortex is a fundamental bit. Therefore, a vortex ``erased" from the sample by the transverse Nernst current takes away exactly the entropy $k_{\rm B}{\rm ln}\,2$ according to the Landauer bound formula\,\cite{landauer}.

% If authors have biography, please use the format below
%\section*{Short Biography of Authors}
%\bio
%{\raisebox{-0.35cm}{\includegraphics[width=3.5cm,height=5.3cm,clip,keepaspectratio]{Definitions/author1.pdf}}}
%{\textbf{Firstname Lastname} Biography of first author}
%
%\bio
%{\raisebox{-0.35cm}{\includegraphics[width=3.5cm,height=5.3cm,clip,keepaspectratio]{Definitions/author2.jpg}}}
%{\textbf{Firstname Lastname} Biography of second author}

% The following MDPI journals use author-date citation: Admsci,  Arts, Econometrics, Economies, Genealogy, Humanities, IJFS, Jintelligence, JRFM, Languages, Laws, Literature, Religions, Risks, Social Sciences. For those journals, please follow the formatting guidelines on http://www.mdpi.com/authors/references
% To cite two works by the same author: \citeauthor{ref-journal-1a} (\citeyear{ref-journal-1a}, \citeyear{ref-journal-1b}). This produces: Whittaker (1967, 1975)
% To cite two works by the same author with specific pages: \citeauthor{ref-journal-3a} (\citeyear{ref-journal-3a}, p. 328; \citeyear{ref-journal-3b}, p.475). This produces: Wong (1999, p. 328; 2000, p. 475)

%%%%%%%%%%%%%%%%%%%%%%%%%%%%%%%%%%%%%%%%%%
%% for journal Sci
%\reviewreports{\\
%Reviewer 1 comments and authors’ response\\
%Reviewer 2 comments and authors’ response\\
%Reviewer 3 comments and authors’ response
%}
%%%%%%%%%%%%%%%%%%%%%%%%%%%%%%%%%%%%%%%%%%

\begin{thebibliography}{999}
%\bibitem[Author1(year)]{ref-journal}
%Author~1, T. The title of the cited article. {\em Journal Abbreviation} {\bf 2008}, {\em 10}, 142--149.



\bibitem{nernst} Behnia,\,K.;  Aubin,\,H. Nernst effect in metals and superconductors: a review of concepts and experiments. {\it 
	Rep. Progr. Phys.}\,{\bf 2016}, {\em 79}, 046502.  
%\bibitem{nernst} Behnia,\,K. and Aubin,\,H. Nernst effect in metals and superconductors: a review of concepts and experiments. {\it Rep. Progr. Phys.}\,{\bf 79}, 046502 (2016). 

\bibitem{varl2018}
Varlamov,\,A.A.; Galda,\,A.; Glatz,\,A. Fluctuation spectroscopy: From Rayleigh-Jeans waves to Abrikosov vortex clusters. {\it
	Rev. Mod. Phys}.\,{\bf 2018},\,{\em 90}, 015009.
%\bibitem{varl2018}
%Varlamov,\,A.\,A., Galda,\,A.\& Glatz,\,A. Fluctuation spectroscopy: From Rayleigh-Jeans waves to Abrikosov vortex clusters. \textit{
%	Rev. Mod. Phys}.\,\textbf{90}, 015009 (2018).

%\bibitem{varl2020} 
%A.\,Glatz , A.\,Pourret, and A.\,A.\,Varlamov, Analysis of the ghost and mirror fields in the Nernst signal induced by superconducting fluctuations. \textit{Phys. Rev}.\,B\textbf{102}, 174507 (2020).

%\bibitem{Matsuda}
%Yamashita,\,T., Shimoyama,\,Y.,  Haga,\,Y., Matsuda,\,T.,  Yamamoto,\,E., Onuki,\,Y., Sumiyoshi,\,H., Fujimoto,\,S., Levchenko,\,A., Shibauchi,\,T. \& Matsuda,\,Y. Colossal thermomagnetic response in the exotic superconductor URu$_2$Si$_2$. \textit{Nat. Phys}.\,\textbf{11}, 17 -- 20 (2015).

\bibitem{chang} Chang,\,J.; Doiron-Leyraud,\,N.; Cyr-Choini\`ere,\,O.; Grissonnanche,\,G.; Lalibert\'e,\,F.; Hassinger,\,E.; Reid,\,J.-P.; Daou,\,R.; Pyon,\, S.; Takayama,\, T.; et al. Decrease of upper critical field with underdoping in cuprate superconductors. {\it Nat. Phys.} {\bf 2012},\,{\em 8}, 751--756. 
%\bibitem{chang} Chang,\,J., Doiron-Leyraud,\,N., Cyr-Choini\`ere,\,O., Grissonnanche,\,G., Lalibert\'e,\,F., Hassinger,\,E., Reid,\,J-Ph., Daou,\,R., Pyon,\, S., Takayama,\, T., Tagaki,\, H., \& Taillefer,\, L., Decrease of upper critical field with underdoping in cuprate superconductors, {\it Nature Physics} {\bf 8} 751-756 (2012). 


\bibitem{universal} 
Rischau,\,C.W.; Li,\,Y.; Fauqu\'e,\,B.; Inoue,\,H.; Kim,\,M.; Bell,\,C.; Hwang,\,H.Y.;  Kapitulnik,\,A.; Behnia,\,K. Universal bound to the amplitude of the vortex Nernst signal in superconductors. {\it 
	Phys. Rev. Lett}.\,{\bf 2021}, {\em 126}, 077001. 
%\bibitem{universal} 
%Rischau,\,C.\,W., Li,\,Y. Fauqu\'e,\,B., Inoue,\,H., Kim,\,M., Bell,\,C., Hwang,\,H.\,Y.,  Kapitulnik,\,A., \& Behnia,\,K. Universal bound to the amplitude of the vortex Nernst signal in superconductors. {\it 
%	Phys. Rev. Lett}.\,{\bf 126}, 077001 (2021). 

\bibitem{huebener}Huebener,\,R.\,P., Ri,\,H.-C., Vortex transport entropy in cuprate superconductors and Boltzmann constant, arXiv:2106.15333. 


\bibitem{bakas1} 
Bakas,\,I. The large-N limit of extended conformal symmetries.  {\it Phys. Lett. B}\,{\bf 1989}, {\em 228}, 57--63. 
%\bibitem{bakas1} 
%Bakas,\,I. The large-N limit of extended conformal symmetries.  {\it Phys. Lett}.\,B\,{\bf 228}, 57 (1989). 

\bibitem{bakas2}
Bakas,\, I. The structure of the $W_{\infty}$ algebra. {\it  Comm. Math. Phys}.\,{\bf 1990}, {\em 134}, 487--508.

\bibitem{pope} 
Pope,\,C.N.; Shen,\,X.;  Romans,\,J.L. $W_{\infty}$ and the Racah-Wigner algebra. {\it Nucl.\,Phys. B}\,{\bf 1990},\,{\em 339}, 191--221. 
%\bibitem{pope} 
%Pope,\,C.\,N., Shen,\,\,X. \& Romans,\,J.\,L. $W_{\infty}$ and the Racah-Wigner algebra. {\it Nucl.\,Phys.}\,B\,{\bf 339}\,(1990). 

\bibitem{shen}
Shen,\,X. $W_{\infty}$ and string theory. {\it Int. J. Mod. Phys. A}\,{\bf 1990}, {\em 7}, 6953 -- 6993.

\bibitem{difrancesco} 
Francesco,\,P.; Mathieu,\, P.;  S\`en\`echal,\,D. {\it Conformal Field Theory}; Springer: New York, NY, USA, 1997. 
%\bibitem{difrancesco} 
%Francesco,\,P., Mathieu,\, P. \& S\`en\`echal,\,D. Conformal field theory. Springer-Verlag, New York (1997). 

\bibitem{ctz1}
Cappelli,\,A.; Trugenberger,\,C.A.; Zemba,\,G. Infinite symmetry in the quantum Hall effect. {\it Nucl. Phys. B}\, {\bf 1993}, {\em 396}, 465--490. 
%\bibitem{ctz1}
%Cappelli,\,A., Trugenberger,\,C.\,A. \& Zemba,\,G. Infinite symmetry in the quantum Hall effect. {\it Nucl. Phys}.\, B\,{\bf 396}, 465 (1993). 

\bibitem{ctz2} 
Cappelli,\,A.; Trugenberger,\,C.A.;  Zemba,\,G. Classification of quantum Hall universality classes by $W_{1+\infty}$ symmetry. {\it 
	Phys. Rev. Lett.}\,{\bf 1994}, {\em 72}, 1902--1905. 
%\bibitem{ctz2} 
%Cappelli,\,A., Trugenberger,\,C.\,A. \& Zemba,\,G. Classification of quantum Hall universality classes by $W_{1+\infty}$ symmetry. {\it 
%	Phys. Rev. Lett}.\,{\bf 71}, 1969 (1993). 

\bibitem{ctz3}
Cappelli,\,A.; Trugenberger,\,C.A.; Zemba,\,G. Stable hierarchical quantum Hall fluids as $W_{1+\infty}$ minimal models. {\it 
	Nucl. Phys.\,B}\,{\bf 1995}, {\em 448}, 470--504. 
%\bibitem{ctz3}
%Cappelli,\,A., Trugenberger,\,C.\,A. \& Zemba,\,G. Stable hierarchical quantum Hall fluids as $W_{1+\infty}$ minimal models. {\it 
%	Nucl. Phys}.\,B\,{\bf 448}, 470 (1995). 

\bibitem{ctsuper}
Trugenberger,\,C.A. 2D superconductivity: classification of universality classes by infinite symmetry. {\it Nucl. Phys.\,B}\,{\bf 2005}, {\em 716}, 509--518.
%\bibitem{ctsuper}Trugenberger,\,C.\,A. 2D superconductivity: classification of universality classes by infinite symmetry. {\it Nucl. Phys}.\,B\,{\bf 716}, 509-518 (2005). 

%\bibitem{kac}
%Kac,\,V. \& Radul,\,A. Quasifinite highest weight modules over the Lie algebra of differential operators on the circle.  {\it
%Comm. Math. Phys.}\,{\bf 157}, 429 (1993). 

\bibitem{ds} 
Diamantini\,M.C.; Sodano,\,P. Characterizing topological order in superconducting systems. \textit{Phys.\,Rev.\,B}\,{\bf 2010}, {\em 82},\,144515. 
%\bibitem{ds} 
%Diamantini\,M.\,C. \& Sodano,\,P. Characterizing topological order in superconducting systems. \textit{Phys.\,Rev}.\,B\,{\bf 82},\,144515\,(2010). 

\bibitem{landauer} Landauer,\,R. Irreversibility and heat generation in the computing process. {\it 
	IBM J. Res. Develop.} {\bf 1961}, {\em 5}, 183. 	
%\bibitem{landauer} Landauer,\,R. Irreversibility and heat generation in the computing process, {\it IBM J. Res. Develop}. {\bf 5}, 183 (1961). 	
	
	
\end{thebibliography}
\end{document}